\documentclass[12pt,eqsecnum,preprint]{aastex}
\begin{document}
\title{PKS 1622-253: A Weakly Accreting, Powerful Gamma Ray Source}
\author{Brian Punsly} \affil{4014 Emerald Street No.116,
Torrance CA, USA 90503 and International Center for Relativistic
Astrophysics, I.C.R.A.,University of Rome La Sapienza, I-00185
Roma, Italy} \email{brian.m.punsly@L-3com.com or
brian.punsly@gte.net}\author{Luis F. Rodriguez} \affil{Centro de
Radioastronomia y Astrofisica, UNAM, Apdo. Postal 3-72 (Xangari),
58089 Morelia, Michoacan, Mexico}
\email{l.rodriguez@astrosmo.unam.mx}\author{Steven Tingay}
\affil{Centre for Astrophysics and Supercomputing, Swinburne
  University of Technology, P.O. Box 218, Hawthorn, Vic 3122, Australia}
\email{stingay@astro.swin.edu.au} \and\author {Sergio Cellone}
\affil{Facultad de Ciencias Astron\'omicas y Geof\'{\i}sicas,
Paseo del Bosque, B1900FWA La Plata, Argentina}
\email{scellone@fcaglp.unlp.edu.ar}
\begin{abstract}
In this Letter, we discuss new deep radio observations of PKS
1622-253 and their implications for the energetics of the central
engine that powers this strong high energy gamma-ray source.
Combining archival infrared and optical measurements with new
millimeter observations, we show that even though the accretion
flow in PKS 1622-253 is under-luminous by quasar standards, a
powerful super-luminal jet is launched with a higher kinetic
luminosity than most EGRET blazars. Only a few percent of the
total jet kinetic luminosity is required to power even the most
powerful gamma ray flares that are observed. The implication is
that a high accretion system is not required to power the
strongest high energy gamma ray sources.
\end{abstract}


\section{Introduction}The connection between accretion flow
parameters and the gamma-ray luminosity from an associated radio
jet is not well understood. Quasars are distinguished by the
strong thermal UV luminosity associated with their accretion flows
and some in the blazar subcategory have been found to have strong
apparent gamma-ray luminosities. Thus, it is of interest to
understand why some of the stronger extragalactic EGRET-identified
sources have accretion flow luminosities below those associated
with quasars (such as the FRII BL-Lac, 0235+164, which has very
weak optical emission in quiescent states, see \citet{tak98}). The
discovery of a sizeable class of strong gamma-ray sources with
accretion flow luminosities less than typical for quasars would
suggest that accretion flow luminosity is, at best, a secondary
(indirect) physical factor in the regulation of gamma-ray emission
from jets.

\par Motivated by this possibility, a study of the not so well
known EGRET source, PKS 1622$-$253, was initiated. Before the
EGRET observations this source was very rarely observed. However,
it is one of the original EGRET detections and has been a firm
identification in each of the three EGRET catalogs
\citep{fic94,har99}. Part of the difficulty in observing this
source is that it lies behind the dense cloud just south of $\rho$
Oph \citep{hun94}. Hence, it is extremely faint in the optical
(visual magnitude, $m_{v}= 22.1$ in NED at z=0.786). Or primary
findings are that only a small percentage of the jet kinetic
luminosity is required to power the observed gamma ray luminosity
and the accretion flow is under-luminous by quasar standards.
\begin{figure}
    \begin{center}
        \includegraphics[scale=0.75]{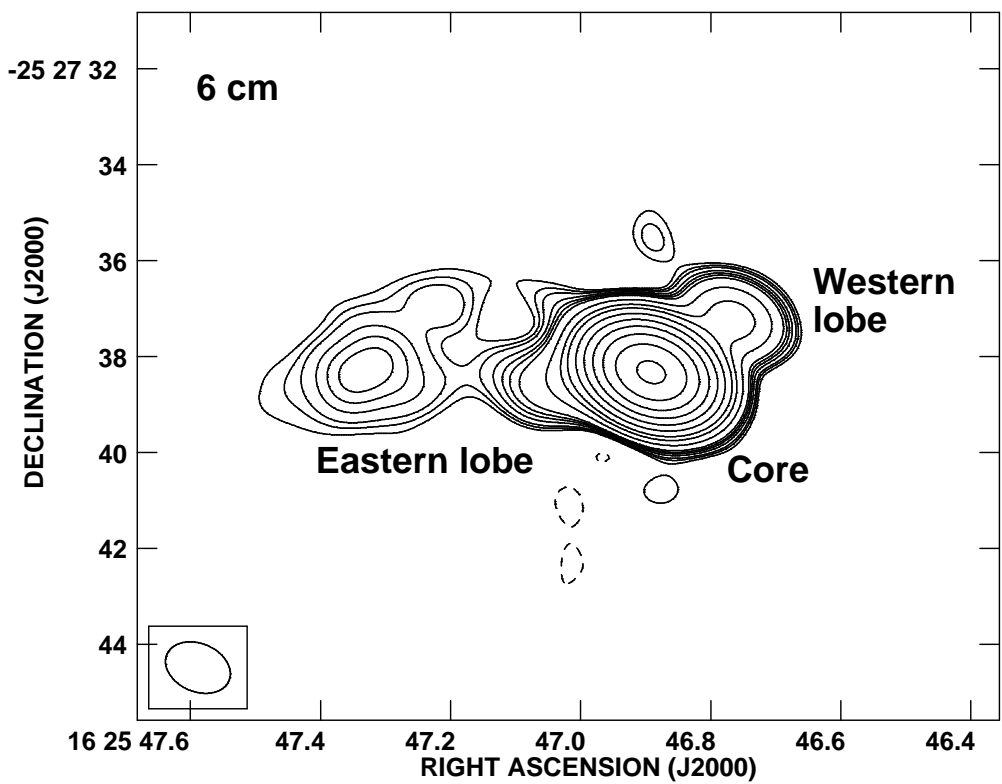}
    \end{center}
    \caption{The 6 cm image of PKS 1622-253. Note the powerful radio core that
    dominates the radio emission. There is a diffuse jet like extension to the east and
    a very strong knot to the northwest that is apparently part of a jet that is
    a continuation
    of the parsec scale jet that is directed at a similar position angle. The contours
    are -4, 4, 5, 6, 8, 10, 12, 15, 30,
60, 100, 200, 400, 800, 1600, 3200, and 6400 times 0.25 mJy
beam$^{-1}$. The peak brightness of the 6 cm image is 1.84 Jy
beam$^{-1}$. The image is restored with a beam dimension of
$1\rlap.{''}40 \times 1\rlap.{''}0$ and position angle of
$67^{\circ}$. The main components of the source are labelled.}
    \end{figure}
\section{The Radio Observations}New observations of PKS 1622$-$253 at 6 and 3.6 cm were made
during June 03, 2002 with the VLA of the NRAO\footnote{The
National Radio Astronomy Observatory is operated by Associated
Universities Inc. under cooperative agreement with the National
Science Foundation.} in the BnA configuration. The data were
reduced using the standard VLA procedures and were self-calibrated
in phase and amplitude, as part of the imaging process. To obtain
images at 6 and 3.6 cm of comparable angular resolution, the 6
cm image was made with the ROBUST parameter of IMAGR set to $-$5,
while the 3.6 cm image was made with the ROBUST parameter of IMAGR
set to 5 and tapered in the (u,v) plane to produce an angular
resolution comparable to that of the 6 cm image. Finally, both
images were restored with the same Gaussian beam. The morphology
of the images were similar and we display the 6 cm radio map in
Figure 1. To obtain the flux densities of the western and eastern
lobes (indicated in Figure 1), we subtracted a point source with
the flux density and position of the core (also indicated in
Figure 1) from the (u,v) data.  Parameters of the core and lobes
measured from the data are listed in Table 1.  The two point
spectral index (calculated from 4.86 GHz to 8.4 GHz) of the core
clearly identifies it as distinct from the characteristically
steep spectrum lobes.
\begin{table}
\begin{center}
\begin{tabular}{ccccc} \hline \hline
Component&19 cm &  6 cm & 3.6 cm & Spectral\\
         & (mJy)& (mJy) & (mJy) &Index\\ \hline \hline
Core     & $1980 \pm 50 $ & $1840\pm 10$ & $1910\pm 10$ & $-0.06\pm 0.01$    \\
West      & $83 \pm 8 $ & $35.6\pm 0.9$ & $24.4\pm 0.7$ & $0.64\pm 0.07$   \\
East     & $145 \pm 15$ &$52.3\pm 1.2$ &$35.1\pm 1.1$ & $0.66\pm
0.07$\\\hline \hline
\end{tabular}\\
\caption{Data reduction from the VLA observations}
\end{center}
\end{table}
To estimate the flux densities of the lobes at a lower frequency,
we used VLA archive data taken at 1.56 GHz on 1991 October 13 in
the A configuration. These results, shown also in Table 1, are not
as reliable as those obtained at 6 and 3.6 cm since the lobes are
only barely resolved from the core.

\par Some limited low frequency radio data are available.  PKS 1622$-$253
appears in the Molonglo Reference Catalog \citep{lar81} with a flux density of 2.36 Jy at 408 MHz and
in the Texas Survey of \citet{dou96} with a flux density of 2.352$\pm$0.047 Jy at 365 MHz.
While the optically thin lobe emission cannot be distinguished from the core at $\sim$400 MHz, an extrapolation
of the VLA data shows that $\sim$0.4 Jy of lobe emission is present at this frequency (assuming that the core had
approximately the same strength when the Molonglo and Texas observations were made).
\section{Estimating the Jet Kinetic Luminosity}
These maps can be used to determine the kinetic luminosity of the
jet from a measurement of an isotropic (not affected by
relativistic beaming) parameter, the strength of lobe emission. We
take this to be the measured data in Table 1.  In a superluminal
source, such as PKS 1622-253, one must be extra cautious about
fluxes being greatly exaggerated by Doppler beaming, hence the
need for an isotropic estimator such as optically thin lobe flux
\citep{pun05}. In order to avoid the ambiguities associated with
Doppler enhancement, we estimate the jet kinetic luminosity from
the isotropic extended emission, applying a method that allows one
to convert 151 MHz flux densities, $F_{151}$, measured in Jy, into
estimates of kinetic luminosity, $Q$, from \citet{wil99,blu00}.
This estimator was expressed explicitly in terms of flux density
in (1.1) of \citet{pun05}:
\begin{mathletters}
\begin{eqnarray}
&& Q \approx 1.1\times
10^{45}\left[(1+z)^{1+\alpha}Z^{2}F_{151}\right]^{\frac{6}{7}}\mathrm{ergs/sec}\;,\\
&& Z \equiv 3.31-(3.65) \nonumber \\
&&\times\left(\left[(1+z)^{4}-0.203(1+z)^{3}+0.749(1+z)^{2}
+0.444(1+z)+0.205\right]^{-0.125}\right)\;,
\end{eqnarray}
\end{mathletters}
where $F_{151}$ is the total optically thin flux density from the
lobes (i.e., no contribution from Doppler boosted jets or radio
cores). We assume a cosmology with $H_{0}$=70 km/s/Mpc,
$\Omega_{\Lambda}=0.7$ and $\Omega_{m}=0.3$.  In order to
implement this technique, one needs to determine which components
are optically thin and which are Doppler enhanced.
\par One might suspect that the strong knot in the western component of
Figure 1 is Doppler enhanced since the parsec scale jet is
directed towards this feature with an apparent velocity of 14c
\citep{jor01,tin98}. However, the integrated flux of the western
component is far weaker than the eastern component and it is
optically thin (see Table 1). Thus, we conclude that the western
component is not likely to be significantly Doppler enhanced. We
can fit the lobe flux densities in table 1 by a two component
model, a strong knot with a spectral index, $\alpha=0.58$, and a
diffuse component with $\alpha=1.0$ from 1.56 GHz to 8.46 GHz. The
western (eastern) component is fit with 10 mJy (20 mJy) of diffuse
emission and 25.6 mJy (32.3 mJy) of knot emission at 4.86 GHz.
This reproduces all of the average flux densities in table 1
except for the 1.56 GHz flux density, 125 mJy, of the eastern lobe
which falls short by 20 mJy. Extrapolating these power laws to 151
MHz, predicts 1.4 Jy of lobe emission and from (3.1), $Q\approx 1.7\times
10^{45}\mathrm{ergs/s}$. If one only considers the diffuse
emission as significant at 151 MHz, one can get a conservative
lower bound of 966 mJy, or $Q>1.2\times 10^{45}\mathrm{ergs/s}$.
\par For the sake of comparison, one can also use the isotropic
estimator derived in (3.8) of \citet{pun05} which can be trivially
adapted to 4.86 GHz,
\begin{eqnarray}
&& Q\approx
\frac{\left[y_{1}(n)\right]^{\frac{n-1}{2}}(486)^{\alpha}}{(n-2)a(n)}10^{42}(1+z)^{1+\alpha}Z^{2}F_{4860}\,\mathrm{ergs/sec}+L\;,
\end{eqnarray}
where the spectral index $\alpha=(n-1)/2$,
$L(\nu)\sim\nu^{-\alpha}$, $F_{4860}$ is the optically thin flux
density at 4.86 GHz in Jy and the total radio luminosity is
$L\equiv\int L(\nu)\,d\nu$. Table 1 indicates that the spectral
index from 4.86 GHz to 1.56 GHz is $\alpha=0.83$. One can get
$a(n)$ and $y_{1}(n)$ from \citet{gin79}. Inserting these values
into (3.8) with the aid of (3.1b) and table 1, we find $Q\approx
1.29\times 10^{45}\mathrm{ergs/s}$. The close agreement between
the two isotropic estimators is reassuring.
\section{Estimating the Accretion Flow Thermal Luminosity}The broad band continuum
spectrum in Figure 2 (after correcting for the Galactic extinction
from the Ophiuchus cloud using the results of \citet{sch98})
indicates a broad synchrotron peak with a high frequency tail with
a spectral index of 1.4 that extends down to $2350\AA$ in the
quasar rest frame \citep{ali94}. Consistent with this
interpretation is the significant polarization in the rest frame
near UV, $2.8\%$ \citep{imp90}.
\begin{figure}
    \begin{center}
        \includegraphics[scale=0.8]{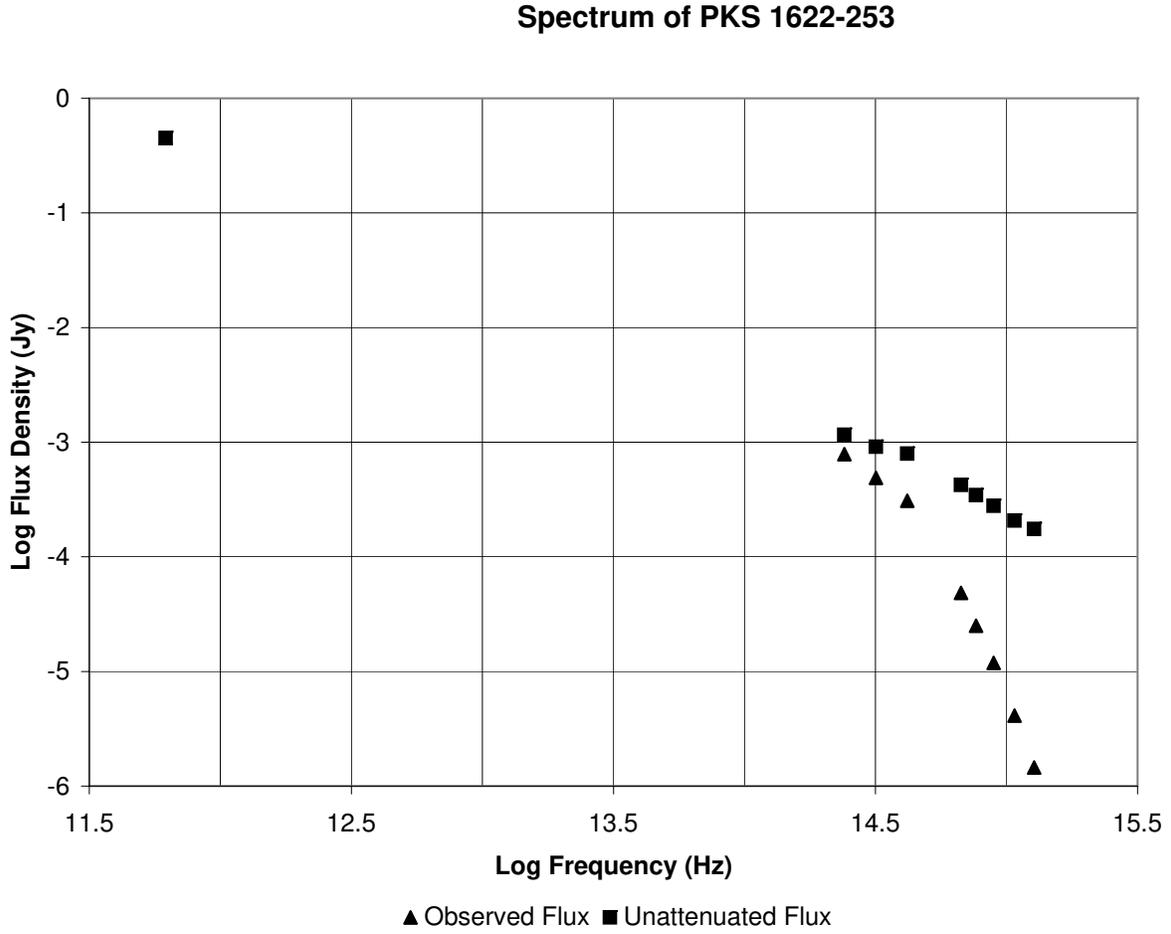}
    \end{center}
    \caption{The continuum broadband spectrum of PKS 1622-253. The spectrum is not
    quasi-simultaneous. The lowest frequency point is
    a millimeter observation courtesy of Ian Robson, the three IR data points
    are from the 2MASS survey and the optical is from \citep{ali94}. The triangles are
    the flux density as directly observed and the squares are corrected for
    Galactic extinction. The frequency is expressed in the rest frame of the radio source.}
    \end{figure}
Thus, one cannot detect the thermal signatures of the accretion
flow. The best one can do is to get an estimate on the upper bound
of the accretion disk luminosity, $L_{bol}$. We tried to improve
on this estimate by making our own optical observations of PKS
1622-253, hoping to catch the source in a low state of blazar
activity and thereby improving the upper bound on the masked
thermal emission. The source was observed on three consecutive
nights with the 2.1-m ``Jorge Sahade'' Telescope, Complejo
Astron\'omico El Leoncito (CASLEO), Argentina, beginning in the
evening of 4/26/03 and ending in the morning of 4/29/03 and was
slightly brighter than that observed by \citet{ali94}, and
therefore was too bright for these observations to improve our
knowledge of the accretion disk luminosity.
\par One can estimate $L_{bol}$ by noting that the bolometric
luminosity of the thermal emission from the accretion flow
includes optical, ultraviolet and any radiation in broad emission
lines from photo-ionized gas or as IR reprocessed by molecular
gas. In order to place an upper bound on $L_{bol}$, we construct a
composite spectral energy distribution (SED) of a quasar accretion
flow. In order to separate the accretion flow thermal luminosity
from IR and optical contamination from the jet, an SED for radio
quiet quasars was chosen since this represents pure accretion
luminosity. In radio quiet quasars, weak versions of radio loud
jets apparently exist \citep{bar04}. The bolometric luminosity of
radio quiet quasars is dominated by IR to UV emission, i.e., the
radio emission from the jet does not contribute significantly to
bolometric luminosity \citep{lao98}. Furthermore, the IR
synchrotron emission from the jet is swamped by thermal dust
emission in radio quiet quasars \citep{haa98,haa00}. Thus, the IR
to X-ray composite of a radio quiet quasar seems to represent the
thermal emission from a viscous accretion flow. A piecewise
collection of power laws is used to approximate the individual
bands of a radio quiet quasar accretion flow luminosity in a
composite spectral energy distribution (SED). The IR and optical
data are from the
      composite spectrum of \cite{elv94}, the NUV (near ultraviolet)
      and EUV data are from the HST composites of \cite{zhe97,tel02}
      and the X-ray portion of the composite is from \cite{lao97}. One can
      compute the bolometric luminosity of the
      accretion flow from the composite spectral energy
      distribution. In addition to the continuum, the
      composite spectrum of \citet{zhe97} indicates that $\approx 25\%$ of the total optical/UV quasar
      luminosity is reprocessed in the broad line region.
      Combining this with the continuum luminosity
      yields, $L_{bol}=1.35\times 10^{46}\mathrm{ergs/sec}$. The absolute visual
      magnitude of the composite is $M_{V}=-25$. Comparing the $2350\AA$ spectral luminosity of PKS 1622-253
      from figure 2, $\nu F_{\nu}< 3\times 10^{45}\mathrm{ergs/sec}$, with the composite
      SED, $\nu F_{\nu}= 4\times 10^{45}\mathrm{ergs/sec}$, one
      finds an upper bound of $L_{bol}< 10^{46}\mathrm{ergs/sec}$ for PKS 1622-253.
      However, since there is no deviation from the steep power law in the spectrum of
      \citet{ali94} at high frequency, we claim that the actual value of $L_{bol}$ from an
      underlying accretion disk is far less than this. One way to
      get a crude estimate of $L_{bol}$ when the blazar jet masks
      the thermal UV flux is to use a broad line luminosity. For
      example, \citet{wan04} use the composite spectrum of
      \citet{fra91} to motivate the relation, $L_{bol}\approx 252.6 L_{H\beta}$,
      where $L_{H\beta}$ is the luminosity of the $H\beta$ broad
      emission line. Using the line strength from \citet{ali94}
      and correcting for Galactic extinction using the results of
      \citet{sch98}, we find $L_{bol}\approx 2.19\times
      10^{45}\mathrm {ergs/s}$.
\par If one also
      assumes that the shape of the composite spectrum is
      independent of quasar luminosity, a simple approximate formula
      is obtained that relates the k-corrected absolute visual magnitude with
      bolometric luminosity,
\begin{eqnarray}
&& L_{bol}\approx 1.35\times10^{\frac{-(25+M_{V})}{2.5}}\times
10^{46}\mathrm{ergs/sec}\;.
\end{eqnarray}
Thus, subtracting out the optical emission from the jet, the
estimate on $L_{bol}$ above suggests that the accretion flow alone
would produce $M_{V}=-23$, right at the typical dividing line
between Seyfert 1s and quasars \citep{ver01}.
\section{Discussion}
\par We performed deep radio observations of PKS 1622-253 with the VLA and
also extricated the intrinsic IR/optical spectrum from the
attenuation due to the Ophiuchus cloud. Our conclusion is that
this source is a classical triple radio galaxy with a powerful
superluminal radio core and powerful FRII radio lobes. The
optical/near UV emission is dominated by the high frequency tail
of a very broad synchrotron spectrum associated with the powerful
radio core. The intrinsic accretion flow luminosity is at the
quasar/ Seyfert 1 dividing line, $M_{V}=-23$ \citep{ver01}. It was
shown that PKS 1622-253 has roughly equal jet kinetic luminosity
and accretion flow bolometric luminosity, $\gtrsim
10^{45}\mathrm{ergs/sec}$. \par The gamma-ray apparent luminosity
has a time average value of $\sim 10^{47}\mathrm{ergs/sec}$ and
flares at $\sim 10^{48}\mathrm{ergs/sec}$ \citep{har99}. The gamma
ray luminosity is enhanced by the Doppler factor to the fourth
power, $\delta^{4}$. In order to determine the intrinsic gamma ray
luminosity of the jets one must know the Doppler factor of the
high energy plasma. The best estimates we have is from
measurements of the apparent super-luminal jet speeds on VLBI
scales. This scale is much larger than the gamma ray emitting
region and it is not clear if one is observing pattern or bulk
flow speeds. Higher resolution typically yields higher apparent
speeds \citep{kel04}. Thus, the maximal apparent super-luminal
velocity from VLBI for each blazar is probably the best estimator
of the Doppler factor of the gamma ray plasma. The super-luminal
motion and the variability results of \citet{kel04} and the
beaming analysis of \citet{pad92} indicate that strong blazar jets
in quasars are typically enhanced by $10^{4}<\delta^{4}<10^{5}$.
The extremely super-luminal jet speed of 14c, \citet{jor01}, and
the extreme radio flux and radio polarization variability,
\citet{tin03}, implies that PKS 1622-253 is an extreme blazar and
$\delta^{4}\gtrsim 10^{4}$ is likely in flare states. Thus, even
during a gamma ray flare $<10\%$ of the intrinsic jet kinetic
luminosity is needed to power the gamma ray emission and on
average $<1\%$ of the intrinsic jet energy is converted into gamma
rays.
\par Finally, we note that such powerful jets from weakly accreting
systems are not rare for EGRET sources. The most extreme case is
PKS 0202$+$149 at z=0.405 with $m_{v}= 22.1$ and a very steep
optical spectral index \citep{per98}. Applying, the methods of
section 4, one concludes $L_{bol}<10^{44}\mathrm{ergs/sec}$. It
does not have FRII radio lobes (or any extended radio emission for
that matter, see \citet{mur93}), so one
 can not make an isotropic estimate of the jet kinetic
luminosity. However, it has been observed more than once with an
apparent gamma-ray luminosity $> 10^{47}\mathrm{ergs/sec}$
\citep{har99}. Apparently, strong jets can be driven without a
strong accretion flow and if there are also large Doppler factors
associated with a nearly line of sight jet, gamma-ray fluxes
detectable by EGRET can be achieved. A plausible theoretical
explanation is provided in \citet{pun01} in which it is claimed
that it is the accumulation of vertical magnetic flux near the
black hole that is the primary determinant of jet power in FRII
radio sources, not the accretion flow proper. The release of black
hole spin energy by the large scale torque applied by the magnetic
flux in the form of a relativistic jet is clearly shown in accord
with the theory in the simulations of \citet{sem04}.

\end{document}